\documentclass[
 aps, pra,
 amsmath,amssymb,
 11pt,
 final,
tightenlines,
 twoside,
 onecolumn,
 nofloats,
nofootinbib,
 superscriptaddress,
showkeys,
showkeywords,
 ]
{revtex4-2}

\usepackage[T2A]{fontenc}
\usepackage[utf8x]{inputenc}
\usepackage[english]{babel}
\usepackage{graphicx}% Include figure files
\usepackage{dcolumn}% Align table columns on decimal point
\usepackage{bm}% bold math
\setcitestyle{authoryear,round,aysep={,},citesep={;}}
\def\squareforqed{\hbox{\rlap{$\sqcap$}$\sqcup$}}

\def\sq{\ifmmode\squareforqed\else{\unskip\nobreak\hfil
\penalty50\hskip1em\null\nobreak\hfil\squareforqed
\parfillskip=0pt\finalhyphendemerits=0\endgraf}\fi}

\def\degr{\hbox{$^\circ$}}

\def\utw{\smash{\rlap{\lower5pt\hbox{$\sim$}}}}

\def\udtw{\smash{\rlap{\lower6pt\hbox{$\approx$}}}}

\def\diameter{{\ifmmode\mathchoice
{\ooalign{\hfil\hbox{$\displaystyle/$}\hfil\crcr
{\hbox{$\displaystyle\mathchar"20D$}}}}
{\ooalign{\hfil\hbox{$\textstyle/$}\hfil\crcr
{\hbox{$\textstyle\mathchar"20D$}}}}
{\ooalign{\hfil\hbox{$\scriptstyle/$}\hfil\crcr
{\hbox{$\scriptstyle\mathchar"20D$}}}}
{\ooalign{\hfil\hbox{$\scriptscriptstyle/$}\hfil\crcr
{\hbox{$\scriptscriptstyle\mathchar"20D$}}}}
\else{\ooalign{\hfil/\hfil\crcr\mathhexbox20D}}%
\fi}}

\newcommand{\aap}{Astron. and Astrophys. }
\newcommand{\aaps}{Astron. and Astrophys. Suppl. }
\renewcommand{\apj}{Astrophys.~J. }
\newcommand{\apjs}{Astrophys.~J. Suppl. }
\newcommand{\mnras}{Monthly Notices Royal Astron. Soc. }
\newcommand{\pasa}{Publ. Astron. Soc. Australia }
\newcommand{\pasj}{Publ. Astron. Soc. Japan }
\newcommand{\pasp}{Publ. Astron. Soc. Pacific }
\renewcommand{\nat}{Nature }

\newcommand{\apjl}{Astrophys.~J.}

\begin{document}

\selectlanguage{english}


\title{The Method of Searching for Rotations of the Polarization Position Angle of Quasars}

\author{\firstname{S.~S.}~\surname{Savchenko}}
\email{s.s.savchenko@spbu.ru, savchenko.s.s@gmail.com}
\affiliation{St. Petersburg University, St. Petersburg, 199034 Russia}
\affiliation{Central (Pulkovo) Astronomical Observatory, Russian Academy of Sciences, St. Petersburg, 196140 Russia}
\affiliation{Special Astrophysical Observatory, Russian Academy of Sciences, Nizhnii Arkhyz, 369167 Russia}

\author{\firstname{D.~A.}~\surname{Morozova}}
\email{d.morozova@spbu.ru, comitcont@gmail.com}
\affiliation{St. Petersburg University, St. Petersburg, 199034 Russia}

\author{\firstname{S.~G.}~\surname{Jorstad}}
\affiliation{St. Petersburg University, St. Petersburg, 199034 Russia}
\affiliation{Institute for Astrophysical Research, Boston University, Boston, MA 02215 USA}
\author{\firstname{D.~A.}~\surname{Blinov}}
\affiliation{Institute of Astrophysics, Foundation Research and Technology-Hellas, Heraklion, GR-71110 Greece}
\affiliation{Department of Physics and Institute for Theoretical and Computational Physics, University of Crete,
Heraklion, GR-70013 Greece}
\author{\firstname{G.~A.}~\surname{Borman}}
\affiliation{Crimean Astrophysical Observatory, Russian Academy of Sciences, Nauchny, 298409 Russia}
\author{\firstname{A.~A.}~\surname{Vasilyev}}
\affiliation{St. Petersburg University, St. Petersburg, 199034 Russia}
\author{\firstname{T.~S.}~\surname{Grishina}}
\affiliation{St. Petersburg University, St. Petersburg, 199034 Russia}
\author{\firstname{A.~V.}~\surname{Zhovtan}}
\affiliation{Crimean Astrophysical Observatory, Russian Academy of Sciences, Nauchny, 298409 Russia}
\author{\firstname{E.~N.}~\surname{Kopatskaya}}
\affiliation{St. Petersburg University, St. Petersburg, 199034 Russia}
\author{\firstname{E.~G.}~\surname{Larionova}}
\affiliation{St. Petersburg University, St. Petersburg, 199034 Russia}
\author{\firstname{I.~S.}~\surname{Troitskiy}}
\affiliation{St. Petersburg University, St. Petersburg, 199034 Russia}

\author{\firstname{Yu.~V.}~\surname{Troitskaya}}
\affiliation{St. Petersburg University, St. Petersburg, 199034 Russia}

\author{\firstname{E.~V.}~\surname{Shishkina}}
\affiliation{St. Petersburg University, St. Petersburg, 199034 Russia}

\author{\firstname{E.~A.}~\surname{Shkodkina}}
\affiliation{St. Petersburg University, St. Petersburg, 199034 Russia}
\begin{abstract}
  Observations of quasars show that the polarization position angle of the emission coming from them varies greatly over time, including periods called rotations during which the angle changes in an orderly manner. The study proposes a method for identifying such events and assessing their statistical significance. The operation of the method is demonstrated using the example of long-term polarimetric observations of the blazars CTA 102, 3C 454.3, and OT 081. During the analysis of light curves, 51 rotations of the polarization position angle were found and it was shown that for CTA 102 and 3C 454.3 the rotations are predominantly oriented in one direction.
\end{abstract}

\maketitle

\section{INTRODUCTION}

Active galactic nuclei (AGN), constituting less than 7\% \citep{Roy1995} of the total number of galaxies in the Universe, have been studied with increasing interest for more than half a century. Studies of these objects, which began in the optical range, have now spread to all ranges available for observation: from radio to ultra-high energies. The properties of active galaxies are most clearly manifested in the subclass of blazars. The reason for the extreme properties of blazars is that their jet is oriented almost directly at the observer, and the jet’s relativistically amplified emission dominates the entire wavelength range. A non-thermal spectrum, high variability in flux density, and high and variable polarization are the distinctive characteristics of blazars in the optical range.

Variability of brightness and polarization can occur on long time scales of the order of weeks, months and years, and short scales of days or even within one day. For the first time, such ultra-fast (intra-day) polarization variability was discovered in 1972 by V.~A.~Hagen-Thorn for an extragalactic object, the blazar OJ\,287, when within one hour a change in the degree of polarization by 2.5\% and a change in the polarization position angle $\chi$ of about 10$^\circ$ were observed \citep{Hagen1972}.

The polarization of optical emission is explained by its synchrotron nature, and the direction of the electric vector position angle (EVPA) is perpendicular to the projection of the magnetic ﬁeld onto the celestial plane. The speciﬁc values of the degree and EVPA depend on the structure of the magnetic ﬁeld in the emitting region and the number of emitting regions along the observer’s line of sight. Typically, ﬂux density and polarization change in a chaotic manner, which is consistent with the random walk model \citep{Moore1982, Marscher2014, Kiehlmann2016}.

However, in some cases, rotations of the EVPA are smooth, long-lasting, and have a large amplitude, which is most often observed during ﬂare activity in a wide range of wavelengths. For the ﬁrst time, the relationship between the EVPA rotations in the optical and radio ranges was discovered for the object OJ~287 in the work of \citet{Kikuchi1988}. Later, in the work of \citet{Marscher2008} a similar behavior for one of the BL~Lac blazar ﬂares was observed using VLBI observations to show that the rotation is associated with the appearance of a new superluminal component passing through the jet core at millimeter wavelengths. Later, in a number of works (for example, \citealt{Marscher2010}) a similar behavior was discovered in other blazars during individual ﬂares. However, not every rotation is accompanied by the appearance of a new component from the jet core at millimeter wavelengths \citep{Jorstad2016}.

Currently, among the attempts made to search and analyze a large number of the EVPA rotations, the RoboPol project \citep{Blinov2019} stands out; this is an instrument and observation program that was designed to systematically study the optical polarization of blazars. As part of the program, regular observations of a selection of blazars were carried out from 2013 to~2017 and 40~rotations were found in 24~objects.

Since the direction of the EVPA is related to the magnetic ﬁeld, a detailed study of the variability of the angle will provide information about the ﬁne structure of blazars jets. Rotations can be generated by both deterministic and stochastic processes. Deterministic processes are associated with ordered magnetic ﬁelds, for example, shock waves traveling along a jet \citep{Marscher2008, Marscher2010}, jet curvature \citep{Nalewajk2010}, and a two-component model \citep{Cohen2020}. Stochastic processes are characterized by entangled magnetic ﬁelds and turbulent plasma motion \citep{Marscher2014,Kiehlmann2017}. Obtaining as large a sample of EVPA rotations as possible will clarify which rotations can be explained by deterministic processes and which are associated with chaotic changes. In addition, since many rotations are observed during ﬂare activity in the gamma-ray range \citep{Marscher2010, Blinov2018}, their study will help provide a better understanding of the physical relationship between optical synchrotron and high-energy emission, and determine the structure of the magnetic ﬁeld in the emitting areas.

Multi-wavelength data analysis during rotation periods became of great interest after the launch of the IXPE (Imaging \mbox{X-ray} Polarimetry Explorer) instrument and the ﬁrst measurements of X-ray polarization of blazars \citep{DiGesu2022}. For example, \cite{DiGesu2023} found a very rapid rotation of the polarization angle in the X-ray range for Mrk\,421 (about 85$^\circ$ per day for ﬁve days), while the optical EVPA remained constant. Such comparisons provide very important information about the location of emitting regions and the structure of the magnetic ﬁeld \citep{DiGesu2023}. The possible connection of astrophysical neutrinos with blazars is currently being actively discussed. At least some of the rotations associated with the repeating structure of gamma-ray bursts \citep{Blinov2021} may in turn be associated with neutrinos \citep{Novikova2023}, so creating a large sample of rotations in the future is important when searching for correlations with high-energy neutrino detection events.

The study of rotations is a challenging task. First, rotations are relatively rare events, therefore, a long series of observations is necessary. Secondly, the measured angle contains ambiguity $\pm\pi n$, the resolution of which imposes even more stringent requirements on the density of the series. Thus, the task of purposefully searching for rotations and identifying them in the light curve is diﬃcult, primarily from an observational point of view.

This work proposes a new method for detecting polarization angle rotations and assessing their reliability, taking into account the experience of similar studies in previous works (for example, \citet{Blinov2015,Blinov2016a,Blinov2016b,Blinov2019,Kiehlmann2016}, etc.).

The structure of the article is as follows. In Section~\ref{sec:observations} we describe the acquisition of observational data. Section~\ref{sec:method} describes the method for ﬁnding rotations. The results of applying this method to observations of three blazars are given in Section~\ref{sec:resuls}. Section~\ref{sec:conclusions} contains conclusions.

\section{OPTICAL OBSERVATIONS}
\label{sec:observations}

The data used in this work was obtained by the authors as part of a monitoring program for a sample of bright gamma-ray blazars, carried out at Saint Petersburg State University\footnote{\url{https://vo.astro.spbu.ru/program/}}. Optical photometric and polarimetric data were obtained in the $R$ band on the following telescopes: LX-200 (40~cm, Saint Petersburg State University, Peterhof), AZT-8 (70~cm, Crimean Astrophysical Observatory, Nauchny), Perkins (1.83~m, Lowell Observatory, Flagstaﬀ, Arizona, USA). The telescopes LX-200 (CCD camera FLI MicroLine ML4710) and \mbox{AZT-8} (CCD camera SBIG ST-7) are equipped with almost identical polarimeters from Saint Petersburg State University. Polarimetric observations were carried out using two Savart plates rotated against each other by 45$^\circ$. The relative Stokes parameters $q$ and $u$ can be obtained from two separate images of each source in the ﬁeld by observing from each plate in turn. The Perkins telescope is equipped with a PRISM\footnote{\url{https://www.bu.edu/prism/}} instrument with a CCD camera and a rotating half-wave plate polarimeter. To determine the polarization, four measurements are made at position angles of 0$^\circ$, 45$^\circ$, 90$^\circ$ and 135$^\circ$.

The polarization measurements were carried out in the $R$ ﬁlter on the AZT-8 and Perkins telescopes; on the LX-200 telescope, the measurements were carried out in “white light” (without a ﬁlter) with a central wavelength $\lambda_{\rm eff}=670$~nm; starting from the fall of 2018 the $R$ ﬁlter was used. Instrumental polarization was determined from stars located near the object, under the assumption that their emission was unpolarized. As a rule, errors do not exceed 1\% for the degree of polarization and 10$^\circ$ for the EVPA for objects with a stellar magnitude of about $17^{\rm m}$.

Details of data acquisition and processing for LX-200 and AZT-8 are given in \citet{Larionov2008}, and for the Perkins telescope in the paper by \citet{Jorstad2010}.

\begin{figure*}
  \includegraphics[width=0.98\textwidth, bb=0 12 425 166,clip]{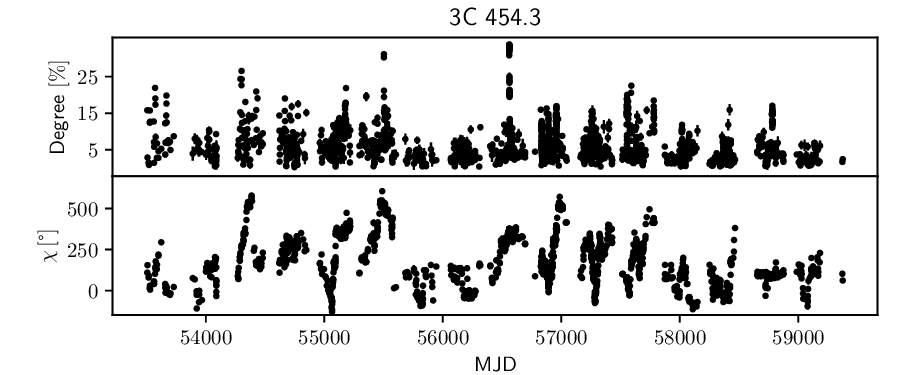}
  \includegraphics[width=0.98\textwidth, bb=0 12 425 166,clip]{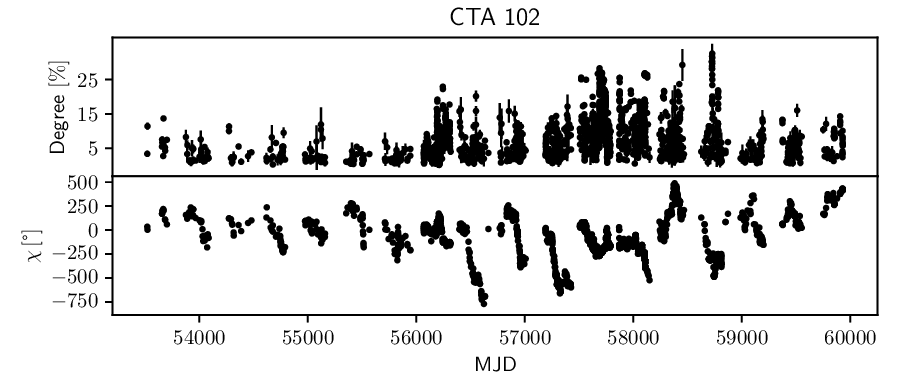}
  \includegraphics[width=0.98\textwidth, bb=0 0 425 166,clip]{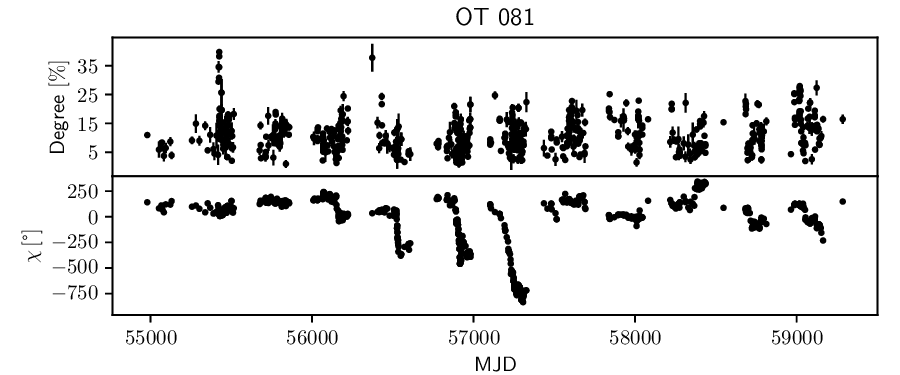}
  \caption{The observational data used in the work are presented from top to bottom: the dependence of the degree and angle of polarization on time for 3C\,454.3 (a), CTA\,102 (b) and OT\,081 (c), respectively. Due to the large range of EVPAs, the error bars for them are smaller than the icon size.}
 \label{fig:curves}
\end{figure*}

The data for three objects, 3C\,454.3 (2005–-2021), CTA\,102 (2005–-2022) and OT\,081 (2009–-2021), was used in this work. Figure~\ref{fig:curves} shows the behavior of the degree and angle of polarization as a function of time for the objects mentioned above.

\section{METHOD FOR ISOLATING ROTATIONS OF POLARIZATION POSITION ANGLE}
\label{sec:method}

In this section, we will describe a technique for searching for signiﬁcant rotations of the polarization position angle in the observational data we obtained. Generally, such events occur suddenly and are unevenly distributed in the curve of the EVPA. The presence of such events in short observation sessions is often determined by eye, by the presence or absence of large-scale trends in the curve of EVPA after which a section of the curve containing rotation is isolated and the angle its are determined. A systematic search program for rotations based on long-term observations must operate with a more stringent criterion that allows such a search to be performed uniformly across the entire light curve to obtain the most complete sample of rotations possible.

For example, in the work of \citet{Blinov2015} the criterion that a section of the EVPA curve contains a rotation is a monotonic and signiﬁcant (exceeding measurement errors) change in the EVPA consistently in at least four observations in a row and with a total amplitude of more than 90$^\circ$. This approach made it possible to detect 14 rotations in the light curves of 12 blazars obtained in the observing season of 2013. A similar approach is used in a number of other works \citep[for example,][]{Blinov2016a, Blinov2016b, Liodakis2017}. The undoubted advantage of this method of searching for rotations is its simplicity and extreme transparency of the results, however, it is not without a very signiﬁcant drawback: the strict requirement of monotonic changes in the EVPA leads to the fact that individual points deviating from monotonic behavior either break one long rotation into several separate episodes, or rotation is not detected at all. Such points deviating from monotonicity can be both a manifestation of measurement noise and a consequence of the presence of several sources of polarized optical emission in the active nucleus: even if one source leads to a long-term and smooth rotation of the polarization vector, short-term bursts of other sources with diﬀerent polarization parameters can lead to the fact that the observed total Stokes parameters exhibit complex behavior and the polarization angle during such ﬂares deviates from monotonic behavior while maintaining the general trend. The authors of the method note this problem and manually combine individual parts of large rotations, broken up by periods of non-monotonicity \citep[for example, Fig. 2 in][]{Blinov2015}.

In this work, we propose an alternative approach to the criterion for identifying the rotation periods of the observed polarization vector. The new criterion should have the following properties:

\begin{list}{}{
\setlength\leftmargin{5mm} \setlength\topsep{0.5mm}
\setlength\parsep{0mm} \setlength\itemsep{0.5mm} }
    \item[1.] Allow individual episodes of non-monotonicity to exist within the rotation to solve the problem of breaking the rotation into separate parts.
    \item[2.] Not be tied to the full range of a rotation. Observations show that relatively long but slow rotations can exist so that the total amplitude is low. Imposing a strict limit on the rotation amplitude can lead to the loss of a certain proportion of such events. Moreover, it introduces a subjective parameter into the measurement process: a restriction on the minimum value of the rotation amplitude.
    \item[3.] It should produce a certain value characterizing the reliability of rotation detection, that is, allowing one to estimate the probability that such a rotation occurred randomly. Even in the absence of a source producing emission with a truly monotonically rotating polarization vector, the variable emission of individual jet cells turbulently varying in time can randomly add up in such a way that the total polarization exhibits a monotonically rotating angle over a fairly long period of time \citep{Marscher2014, Kiehlmann2016}. In long-term observational data, such random rotations can be present in considerable quantities, and there must be a criterion that allows one to cut oﬀ insigniﬁcant rotations.
\end{list}

Before proceeding directly to the description of the criteria, the need for general preprocessing of the EVPA curve should be noted. First, the ambiguity of the position angle $\chi$ must be resolved: since its value is determined with an ambiguity of $\pm\pi n$, the rotations will inevitably contain discontinuities when the angle passes through 0/180~degrees. In practice, the standard approach to solving this problem \citep[for example,][]{Abdo2010,Ikejiri2011,Blinov2015} is the assumption of smooth behavior of the EVPA. In this case, if two neighboring points diﬀer by more than 90\degr, then the value $\pm\pi n$ is added to the second one, where $n$ is selected in such a way as to minimize the diﬀerence. This approach makes it possible to reconstruct long-term rotations with an amplitude of hundreds of degrees, but can only be applied if there is a suﬃciently dense observational series.

\begin{figure}
  \includegraphics[width=0.6\columnwidth, clip=true, trim=0 0.4cm 0 0]{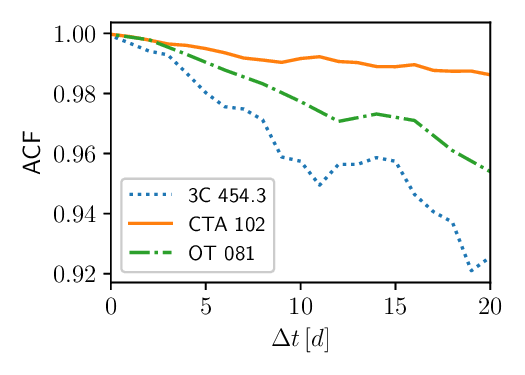}
  \caption{Autocorrelation function for the EVPA of objects 3C\,454.3, CTA\,102 and OT\,081.}
  \label{fig:acf}
\end{figure}

If there is a large gap in the observations, then it can no longer be assumed that there is a correlation between the values at two neighboring points. In this work, we use the approach described above to resolve the ambiguity of the EVPA; however, with large gaps in observations, it cannot be guaranteed that neighboring points are connected, and the EVPA must be counted from zero, and the possible rotation is broken into separate events. The speciﬁc period of time after which EVPAs cannot be considered related depends on the object and its local behavior. The upper limit of this interval can be obtained from the autocorrelation function of the EVPA (it cannot be larger than the interval over which a high correlation between the $\chi$ values remains). Figure~\ref{fig:acf} shows the autocorrelation function calculated using the LDCF method \citep{Welsh1999} for the three sources studied in this work. The values of $\chi$ are strongly correlated even over a fairly long period of time, but it is obvious that the main contribution here comes from states without rotations. In this case, the limitation on the maximum gap in observations will be given by the average rate of an EVPA rotation: if during the observational gap the angle of polarization has time to rotate by more than 180$^\circ$, then it will be impossible to resolve the ambiguity $\pm\pi n$. The characteristic values of the rotation rate of the polarization vector published in the literature are about ten or slightly more degrees per day \citep{Blinov2015, Kiehlmann2016}. Therefore, if there is a gap in observations of about 10–15 days, a complete uncertainty in the angle value will arise. Faster rotations will require denser observations \citep{Kiehlmann2021}.

Another important point is the smoothing of observational data. The relative accuracy of determining polarization parameters is lower than the accuracy of photometric measurements. As a result, the measurement error of the EVPA can signiﬁcantly exceed the measured change of this value between adjacent observations. In this case, only signiﬁcant changes in $\chi$ are important for searching and analyzing rotations. Recently, a data smoothing algorithm based on Bayesian blocks has become widespread \citep{Scargle2013}. The main idea of the method is to replace the observed time series with a piecewise constant function; the time series is divided into non-overlapping intervals (blocks), in which the dependent variable (the position angle $\chi$) is described by a constant value. That is, the partition into blocks is performed in such a way that the measured value within the blocks does not change strongly enough to recognize these changes as signiﬁcant. Individual blocks represent periods between which the value changed signiﬁcantly. Among all possible partitions, one that minimizes the discrepancy between real observations and the approximating piecewise constant function is searched for. The best result will be obtained by a partition in which each observation point is a separate block, but such a partition is meaningless since it does not smooth the data. To solve this problem, an a priori number of blocks that is less than the number of observation points is set; as a result, neighboring points with close $\chi$ values are forced to combine into a single block. The number of partition blocks established a priori is selected from the following considerations: in the absence of a signal, that is, when the EVPA does not change, and the observed changes arise only due to errors, the probability of identifying a random change in a separate false block should be less than 0.05. Note that the chosen a priori number of partition blocks is not the ﬁnal number of blocks into which the time series will be split: it is the expected value of the number of blocks before the splitting begins; in the case of a highly variable signal, the ﬁnal number of blocks will be larger, and in the absence of variability, all points can be combined into one block (this is the Bayesian part of the method: as in any Bayesian modeling, an a priori model is ﬁrst speciﬁed, which changes during the modeling process according to the available observational data). A direct algorithm for ﬁnding the optimal partition is presented in the article by \citet{Scargle2013}. After the partition is completed, the points that fall into the common block are averaged by calculating the weighted average (that is, the errors of individual measurements are taken into account). The advantage of this approach in comparison with other smoothing methods, for example, running average, is the absence of “smearing” of sharp changes in value: sharp changes are always divided into a separate block, which is important in the context of our task.

In the rest of our work, we use the approach described above to smooth the observed EVPA curves. All calculations are performed with smoothed curves, however, for illustrative purposes, we plot both the smoothed curves (in the form of piecewise constant curves) and the original observations (in the form of points) in the ﬁgures containing the dependence of the EVPA on time.

\subsection{Binomial Test}
\label{sec:binomial}

As a basis for a criterion that satisﬁes the stated requirements, we propose to use a one-sided binomial test. In the absence of an ordered direction of rotation of the polarization vector $\chi$ (i.e. when $\chi$ experiences only chaotic changes), both directions of change of the angle between two adjacent measurements are equally probable. In any part of the curve with such a chaotic change in the polarization angle, the number of clockwise changes in the angle should be balanced by the number of counterclockwise changes. If the polarization vector has a dominant direction of rotation, then the number of angle changes in this direction will prevail.

Naturally, measurement errors, randomness, as well as stochastic processes in the jet lead to the fact that in the ﬁrst case (no ordered rotation) the number of changes of $\chi$ in both directions will not be strictly equal whereas in the second case (with rotation) not all the changes will be one-sided. The binomial test will help to distinguish these cases. Let’s assume that $N_{\mathrm{obs}}$ changes of the value of $\chi$ are detected during the observation process. Of these, $N_{\mathrm{cw}}$ occurred clockwise and $N_{\mathrm{ccw}}$ counterclockwise ($N_{\mathrm{obs}} = N_{\mathrm{cw}} + N_{\mathrm{ccw}}$). Also, let’s assume that $N_{\mathrm{ccw}}>N_{\mathrm{cw}}$. The probability that such an imbalance could occur by chance is determined through binomial coeﬃcients by the following equation:

\begin{equation}
  \label{eq:binom_test}
    p_{\mathrm{binom}} = 0.5^{N_{\mathrm{obs}}}\sum_{i\,=\,N_{\mathrm{ccw}}}^{N_{\mathrm{obs}}}\binom{N_{\mathrm{obs}}}{i}.
\end{equation}

The null hypothesis in this case is the assumption that both directions of change of the angle $\chi$ are equally probable (there is no ordered rotation). If the resulting value of $p$ is not close to zero (not lower than a certain signiﬁcance level chosen in advance), then the null hypothesis cannot be rejected and, thus, the presence of a rotation cannot be conﬁrmed. If the value of $p$ turns out to be below a given signiﬁcance level, then in this area there is likely to be an ordered direction of the EVPA rotation.

\subsection{T-Test}
\label{sec:student}

A weakness of the proposed criterion based on the binomial test is its abstraction from the rate of change of the EVPA and from how the average rotation rate relates to the variation of rates between individual measurements. To solve this problem, we propose a second test, based on the requirement of a signiﬁcant average rotation rate.

Let us assume there are $N$ changes in the EVPA between neighboring observations. If in a given section of the light curve variations of $\chi$ are produced only by stochastic processes, without stable rotation, then the average rotation rate will be close to zero. Otherwise, if in addition to stochastic changes, there is also a stable rotation, then the rate averaged over individual measurements will diﬀer from zero, revealing a constant component in the change of the EVPA. Statistical tests (Shapiro\,--\,Wilk and Q\,--\,Q) showed that in certain sections of the EVPA curve, the distribution of the angle variation rates is close to normal, so checking for the signiﬁcance of the diﬀerence of the average rates from zero can be performed using Student’s \mbox{T-test}.

Let us denote by $\chi_i$ and $\chi_j$ the EVPA values measured at times $t_i$ and $t_j$, respectively. Then the EVPA variation rate in this area can be taken as follows:
\vspace{-2mm}
$$ 
r_{ij} = \cfrac{\chi_i-\chi_j}{t_i - t_j}.
$$

If the measurement errors of the polarization angle $\sigma_{\chi_i}$ and $\sigma_{\chi_j}$ are known, then the rate error will be:

\begin{equation} \vspace{-2mm} 
    \sigma_{r_{ij}} = \cfrac{\sqrt{\sigma^2_{\chi_i}+\sigma^2_{\chi_j}}}{t_i - t_j}.
\end{equation}

In this case, the weighted average EVPA variation rate in a certain area will be equal to:
\vspace{-4mm}
\begin{equation}  
    \bar{r} = \cfrac{\sum_{k=1}^N  \cfrac{r_k}{\sigma_{r_k}^2}}{\sum_{k=1}^N \cfrac{1}{\sigma_{r_k}^2}}, \vspace{0mm}
\end{equation}
and its standard deviation:
\begin{equation}
    \sigma_r = \sqrt{\cfrac{\sum_{k=1}^N\cfrac{(r_k-\bar{r})^2}{\sigma_{r_k}^2}}{\cfrac{N-1}{N}\sum_{k=1}^N \cfrac{1}{\sigma_{r_k}^2}}}.
\end{equation}

Having these values, we can calculate the Student’s \mbox{$t$-statistic} to test the null hypothesis that the average EVPA variation rate is equal to zero:
\begin{equation}
    t = \cfrac{\bar{r}}{\sigma_r/\sqrt{N}},
\end{equation}
from where the \mbox{$p$-value} is calculated in a standard way using the Student distribution for a given number of degrees of freedom $\nu=N-1$:

\begin{equation}
    p_{t \hbox{{-}}{\rm test}}(t) = \cfrac{\Gamma\left( \cfrac{\nu+1}{2} \right)}{\sqrt{\nu\pi}\Gamma\left(\cfrac{\nu}{2} \right)} \left( 1 + \cfrac{t^2}{\nu} \right)^{-\frac{\nu+1}{2}}.
\end{equation}
\begin{figure*}
\includegraphics[width=0.49\textwidth]{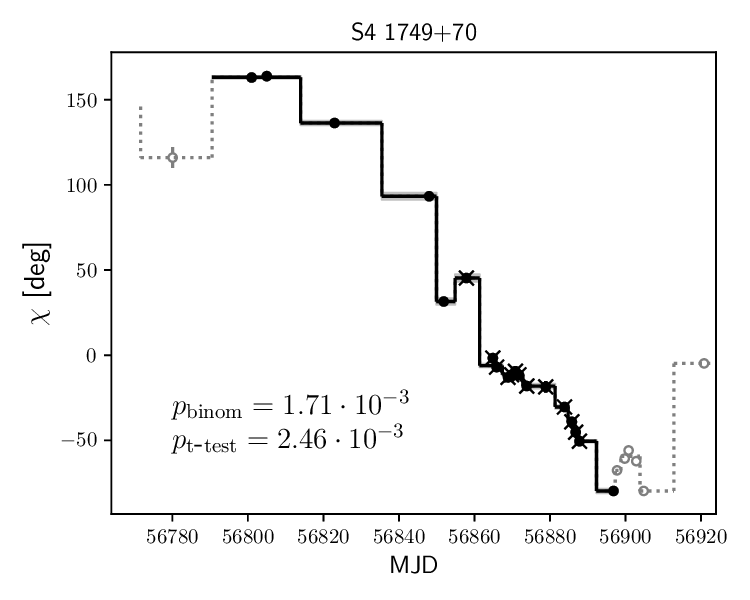}
\includegraphics[width=0.49\textwidth]{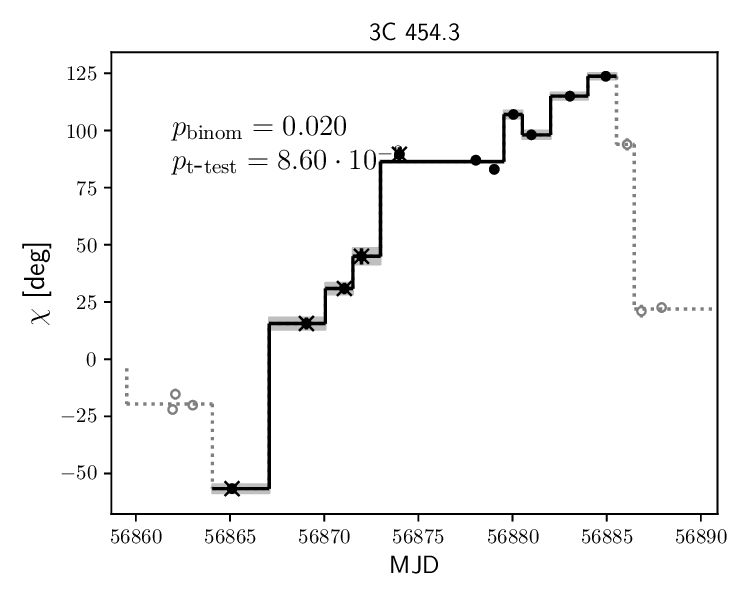}
\caption{
Examples of rotations identiﬁed using the proposed criteria \citep[according to Robopol, ][]{Blinov2020}. Circles with error bars are observational data, lines are data smoothed using Bayesian blocks. Filled circles and a dark continuous line show the area of the selected rotation, open circles and a dotted line show areas outside the rotations. The numbers are the $p$-values of rotations determined by the binomial test ($p_{\mathrm{binom}}$) and the T-test ($p_{t \hbox{{-}}{\rm test}}$). For comparison, crosses indicate points assigned to rotations based on criterion of the strict monotonicity of changes of the EVPA in the work of \cite{Blinov2016b}.}
 \label{fig:robopol_comparison}
\end{figure*}

Values of $p$ close to zero indicate a low probability that the stochastic process will generate such a consistent change in the EVPA and lead to the emergence of an average rate signiﬁcantly exceeding the observed scatter. Otherwise, the average rate of the EVPA in this section does not diﬀer signiﬁcantly from zero, and it cannot be said that a rotation takes place.

\subsection{Discussion of Criteria}
Note that the described criteria meet all the requirements mentioned above:
\begin{list}{}{
\setlength\leftmargin{5mm} \setlength\topsep{1mm}
\setlength\parsep{0mm} \setlength\itemsep{1mm} }
    \item[1)]
 there is no requirement for a strict monotonic change in the polarization angle, as long as the predominance of some speciﬁc direction over the noise component is observed; \item[2)] the full amplitude of a rotation and the average rate do not play any role; if the accuracy of the measurements allows, then with a sfficient number of observations, arbitrarily small and a slow rotation can be detected;
\item[3)] there is a numerical characteristic of a rotation reliability.
\end{list}

Another important point is that the number of observation points inside the rotation naturally aﬀects its signiﬁcance, determined by both criteria. Since the direction of $\chi$ can exhibit a random walk, short periods when EVPA rotates monotonically several times in a row can occur spontaneously, without the presence of mechanisms generating a true rotation. The criterion used in previous works, which requires four or more consecutive one-sided changes in the EVPA to detect a rotation, does not possess this property; as a result, for short segments of the observed curve the probability of false positives increases, since a randomly changing EVPA with a probability of $0.0625$ will rotate four times in a row in one direction (the restriction on the minimum rotation amplitude used in these works only partly solves this problem, since at a low degree of polarization, random walks of the EVPA can be very large \citep{Larionov2016}).

\begin{figure}
  \includegraphics[width=0.49\textwidth]{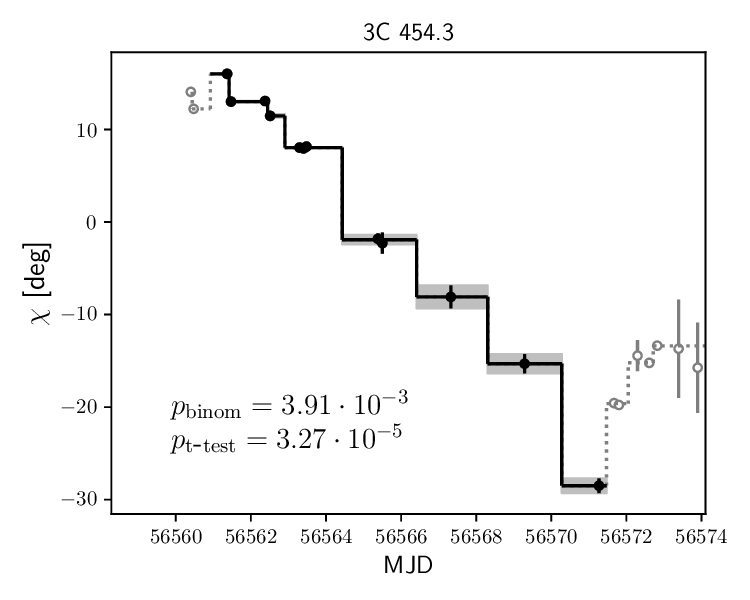}
  \includegraphics[width=0.49\textwidth]{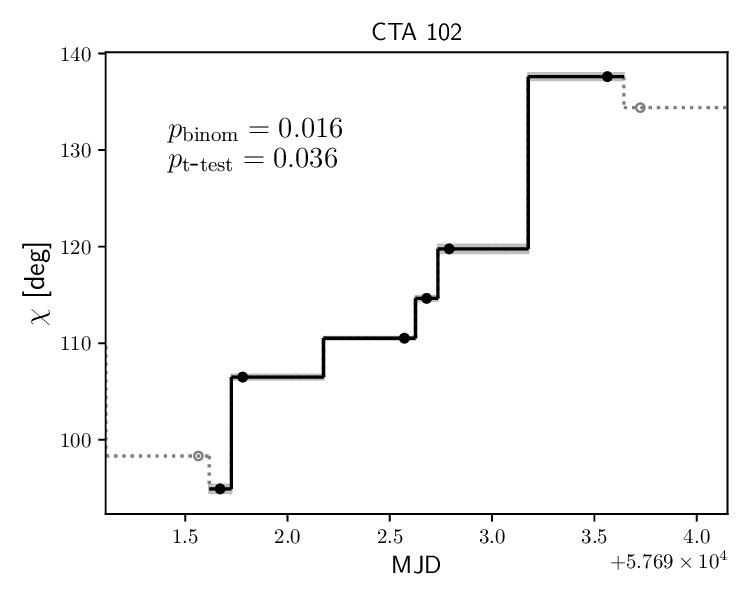}
  \caption{
  Examples of detected rotations with low (less than $90^\circ$) amplitude based on our observational data. The structure of the ﬁgure is similar to Fig.~\ref{fig:robopol_comparison}.}
  \label{fig:low_amplitude_rotations}
\end{figure}

A comparison of the results of identifying rotations using the proposed criteria with a method based on searching for a strictly monotonic EVPA variation is presented in Fig.~\ref{fig:robopol_comparison} via two objects: S4\,1749+70 (left) and 3C\,454.3 (right) according to Robopol data \citep{Blinov2020}. Since in that work a minimum of four one-sided $\chi$ changes in a row was required to detect rotations, which corresponds to a \mbox{$p$-value} of 0.0625 in the binomial test, we used this limit as the signficance level of our tests to detect rotations. The crosses mark the points that were included in the rotations found using the criterion of monotonicity of $\chi$ variation in the work of \cite{Blinov2016b}. Also, since in that work, there was no limit on the maximum gap in observations that would break the rotation, we also did not impose any restrictions for a correct comparison. Our proposed criteria make it possible to identify longer rotations by including new points, separated from the rest by a short period of non-monotonicity. For example, in the case of object S4\,1749+70 (Fig.~\ref{fig:robopol_comparison}a), only a single point (near MJD~58852) deviates from monotonicity, as a result all previous points are lost and are not included in the rotation.

Figure~\ref{fig:low_amplitude_rotations} shows examples of reliable rotations with low amplitude (less than 90\degr) found in observations of objects 3C\,454.3 (Fig.~\ref{fig:low_amplitude_rotations}a) and CTA\,102 (Fig.~\ref{fig:low_amplitude_rotations}b) in our observational data. Although in both cases the EVPA varies monotonically over a signiﬁcant number of observations, the lower bound on the rotation amplitude used in past work to guard against false positives would not allow these rotations to be detected. The calculated $p$-values of both criteria (see Fig.~\ref{fig:low_amplitude_rotations}) show that these rotations are statistically significant.

The ability to detect low-amplitude rotations is especially important because the observed rotation amplitude may be small if there are multiple sources of polarized (and non-polarized) emission in the object. In this case, the observed Stokes vector is the sum of the vectors of individual sources, and the observed direction of the polarization vector will depend not only on the directions of the polarization vectors in these sources but also on their relative intensity and degree of polarization. Thus, if there is a bright constant source of polarized emission in an object and a relatively weak source demonstrating rotation of the angle $\chi$, then the observer, instead of rotating the polarization vector, will observe its oscillations around the preferential direction corresponding to the direction of polarization in the constant source. Moreover, the lower the luminosity and degree of polarization of the variable source, the smaller the oscillation amplitude will be. Our proposed method makes it possible to isolate half-periods of such small oscillations if the measured EVPA error is much smaller than the amplitude of the oscillations; however, an additional check can be attained by the study of structures outlined by the Stokes vector on the $Q-U$ plane (see \citealt{Uemura2016,Shablovinskaya2019})).

\subsection{Numerical Experiments}
\label{sec:numerical}

To test the performance of the proposed criteria, we conducted two numerical experiments, the main goal of which was to demonstrate, on the one hand, the low probability of detecting rotations arising due to random walk of the polarization vector within the limits of measurement errors, and on the other hand, the high probability of detecting a true rotation in noisy data.

For the first experiment, we created artificial curves of the EVPA that do not contain rotations: the true value of $\chi$ is constant for them, but the measurements contain an error, so that the observed values of $\chi$ have a scatter. In practice, the EVPA error increases as the degree of polarization decreases, so that for weakly polarized objects it can be quite large, and random deviations of the measured $\chi$ values can line up in a consistent rotation. If the measurement errors of the EVPA are known, then our proposed method is resistant to such errors, since the calculation of $p_{\mathrm{binom}}$ and $p_{\mathrm{t \hbox{-} test}}$ is preceded by smoothing using Bayesian blocks: random changes in $\chi$ within the errors do not give a significant change between neighboring measurements, so such a walk will be averaged within a single block even if the errors are very large.

However, in practice it may turn out that EVPA errors are underestimated (for example, due to some unaccounted factor). In this case, a random change in $\chi$ can be taken as significant and this measurement will be allocated as a separate block. If the EVPA measurement errors are greatly underestimated, this can lead to the identification of several such blocks, which with some probability can demonstrate a smooth variation of $\chi$. We tested this possibility by introducing an additional factor $f_\sigma$ into the simulation, which determines how much the EVPA error is underestimated:
$$
\sigma_{\chi, \mathrm{used}} = f_\sigma\sigma_{\chi, \mathrm{true}},
$$
where $\sigma_{\chi, \mathrm{true}}$, true is the true EVPA error utilized to create the artificial curves, and $\sigma_{\chi, \mathrm{used}}$, used is the value used in searching for rotations.

Therefore, the numerical experiment looks like this: we created 1000 random EVPA curves, each consisting of 1000 points with a constant value of $\chi$, to which random errors distributed according to the normal law are added. For each curve, a search for rotations was performed, varying the value of $f_\sigma$ in the range from 0.1 to 1.0 (that is, in the extreme case, the measurement error of $\chi$ is underestimated by a factor of 10). It is important to note here that the absolute value of the added error does not matter since the probability that the measured $\chi$ will randomly change in the same direction several times in a row does not depend on the magnitude of the error. Different values of the added error can only change the average rate of such false rotation, but our proposed method is not sensitive to the rate.

\begin{figure}
\includegraphics[width=0.6\textwidth]{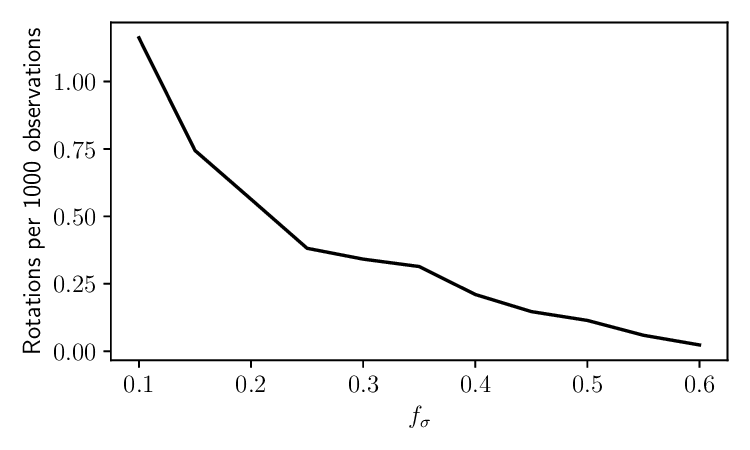}
\caption{Average number of detected random rotations per thousand measurements as a function of underestimation of EVPA measurement error.}
 \label{fig:numerical1}
\end{figure}

Figure~\ref{fig:numerical1} shows the average number of detected random rotations in an EVPA curve consisting of 1000 measurements, depending on how much the errors are underestimated. The figure shows that smoothing observations using Bayesian blocks protects against the detection of random walks of the EVPA within errors (even if the errors are underestimated by half, such false detections are practically excluded). If observational errors are greatly underestimated, then the number of false detections increases.

\begin{figure}
\includegraphics[width=0.6\textwidth, bb=10 25 250 212,clip]{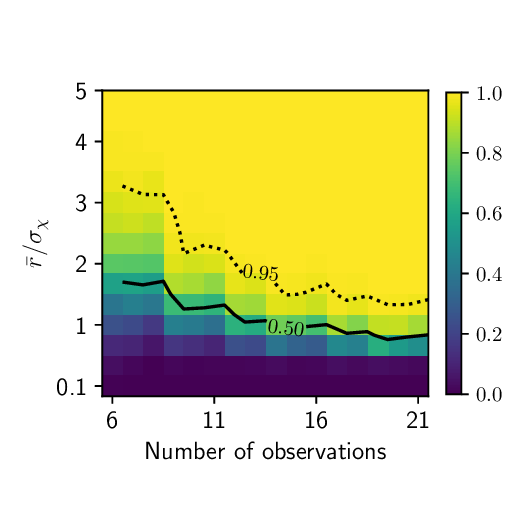}
\caption{The probability of detecting a rotation depending on its duration (number of observations) and the ratio of its average rate to the average observation error.}
 \label{fig:numerical2}
\end{figure}

The second experiment aims to determine the ability to detect true rotations in the presence of large measurement errors. If the errors are large and the rotation rate is low, the systematic change in the EVPA may be less than the scatter of measurements and the rotation may not be detected. It is natural to expect that in our proposed approach this problem can be solved by increasing the number of observations since both proposed criteria are statistical. For example, a binomial test requires significantly more changes of $\chi$ in one direction than in the other. As the measurement error increases, the statistics will deteriorate, since random changes in the EVPA that are opposite to the dominant direction will occur. In any case, on average, there will be more changes in $\chi$ in the dominant direction, and with a sufficient number of observations, it will be possible to accumulate a significant signal. Similarly with the T-test: a sufficiently large number of measurements will allow one to determine that the average change of EVPA is significantly different from zero, even in the presence of errors that exceed the change of EVPA between successive observations.
\begin{figure*}
\includegraphics[width=0.88\textwidth, bb=5 14 415 425,clip]{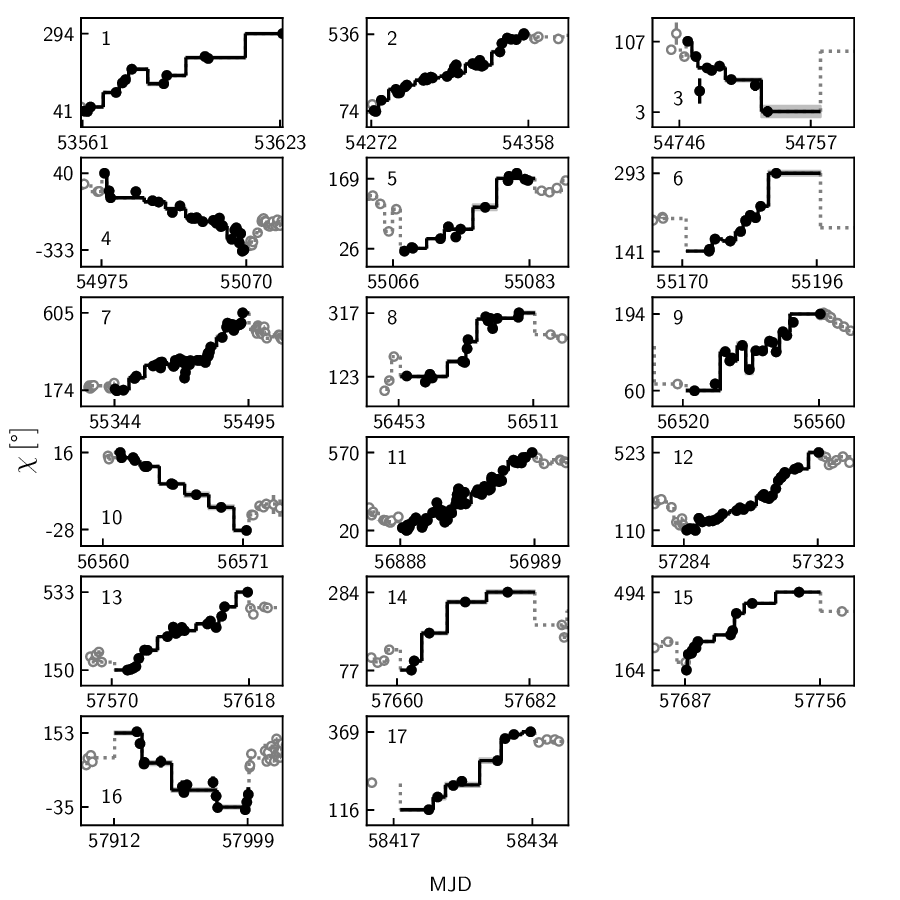}
\caption{All EVPA rotations detected for the object 3C\,454.3.}
 \label{fig:3c454_all_rotations}
\end{figure*}

To demonstrate this, we ran the following simulation. An EVPA curve containing a monotonic variation (i.e. a true rotation) is created; then normally distributed measurement errors are added to the curve. The first parameter of the simulation is the ratio of the average rotation rate to the magnitude of the measurement error (the smaller this ratio, the more difficult it is to detect the rotation since changes in $\chi$ between individual observations begin to “drown” in the measurement errors). The second parameter is the duration of rotation (that is, the number of observations assuming a uniform observational series). 
Next, a search for rotation occurs, and the results are averaged over 1000 implementations for each pair of simulation parameter values. Figure~\ref{fig:numerical2} shows the results of this experiment, demonstrating a limitation of our method: the short, slow rotation is the hardest to detect. For example, if the average daily change in the EVPA is approximately equal to the measurement error, even 14 daily observations in a row will detect such a rotation with only a 50\% probability. To detect a slow rotation reliably, it must be long-lasting so that a significant change in EVPA occurs. If the daily variations of the EVPA are three time larger than the measurement errors, then six observations are almost guaranteed to detect such a rotation.

\section{RESULTS}
\label{sec:resuls}

The criteria described above were applied to search for rotations in observations of three objects: 3C\,454.3 \citep[$z=0.859$,][]{Jackson1991}, CTA\,102 \citep[$z=1.037$,][]{Schmidt1965} and OT\,081 \citep[$z=0.320$,][]{Stickel1993}. These objects, included in the monitoring program of Saint Petersburg State University, were selected based on the presence of long and dense series of observations: the objects are included in the subsample with the highest priority of observations and have periods of regular (almost daily) observations.
\begin{figure*}
\includegraphics[width=0.9\textwidth]{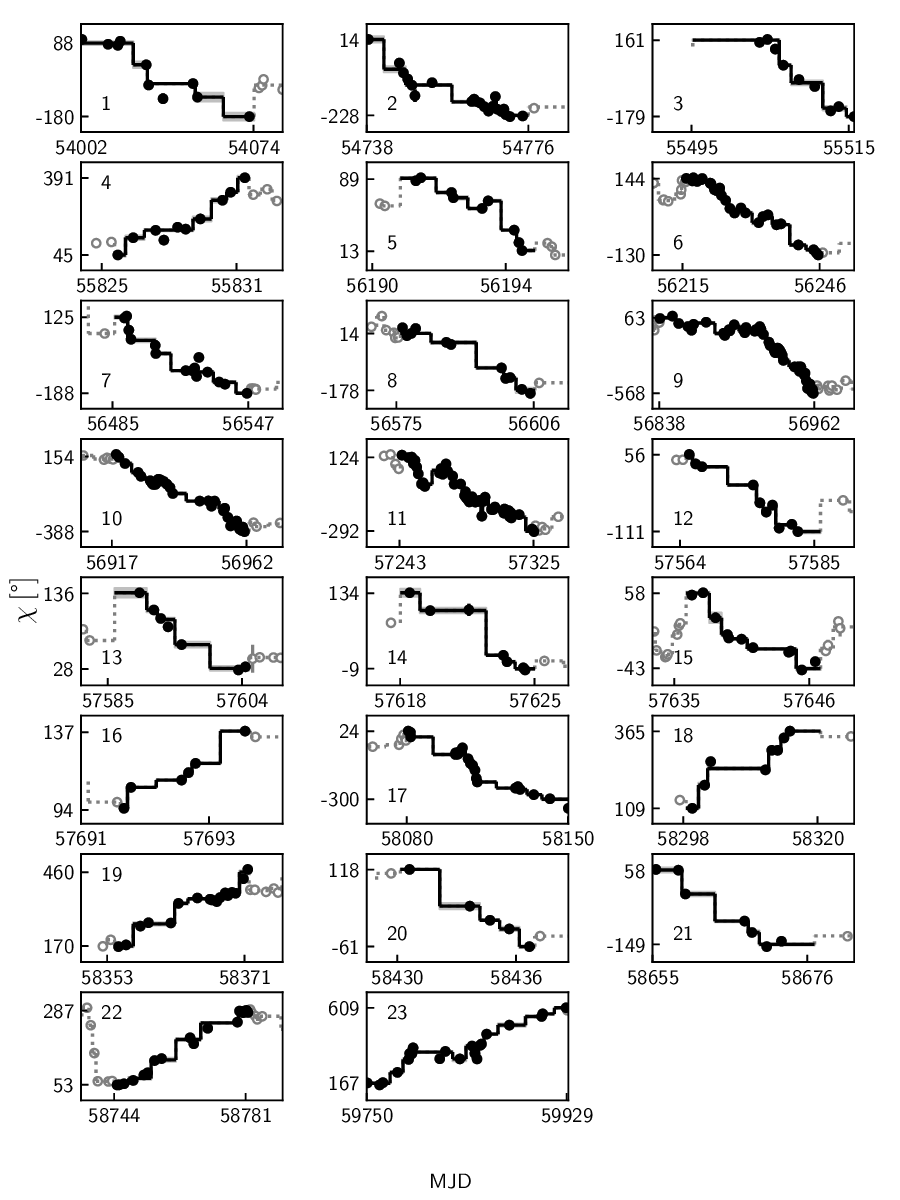}
\caption{All EVPA rotations detected for the object CTA\,102.}
 \label{fig:cta102_all_rotations}
\end{figure*}

\begin{figure*}
\includegraphics[width=0.89\textwidth]{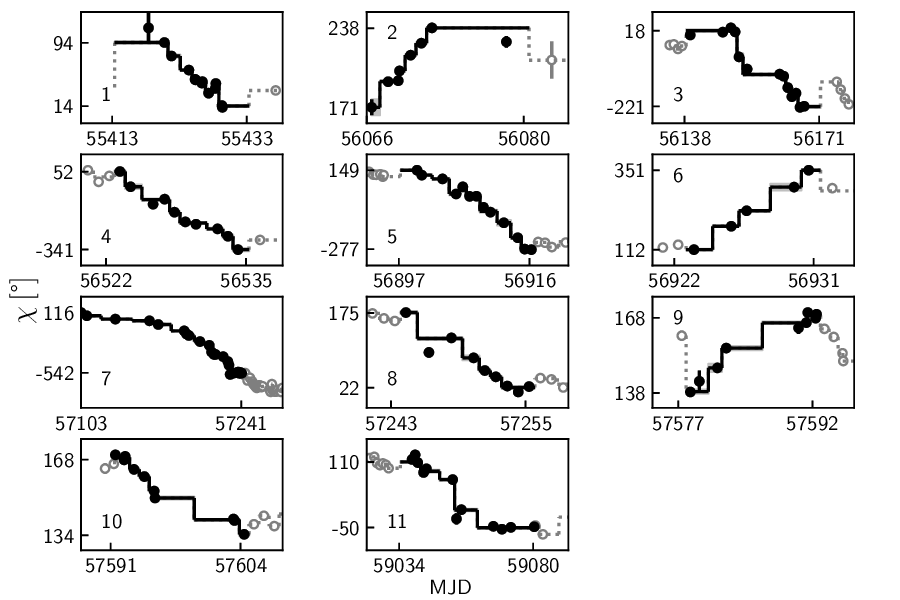}
\caption{All EVPA rotations detected for the object OT\,081.}
 \label{fig:ot081_all_rotations}
\end{figure*}
\renewcommand{\baselinestretch}{0.65}
\setlength{\tabcolsep}{5pt}
\begin{table*}[]
\caption{Rotations parameters of the object 3C\,454.3: date of rotation, $\Delta \mathrm{MJD}$ is the rotation duration, $A$ is the total rotation amplitude (in degrees), $\bar{r}$ is the average rotation rate (in degrees per day), $p_{\mathrm{binom}}$ is the $p$-value by binomial test, and $p_{{t \hbox{-}\rm test}}$ is the $p$-value from T-test}
\label{tab:rotations3C454}
\medskip
\begin{tabular}{c|r|r|r|r@{~}|c|c}
\hline
Number  &  \multicolumn{1}{c|}{MJD} & \multicolumn{1}{c|}{$\Delta \mathrm{MJD}$}  &  $A$,  deg  & $\bar{r}$,  deg day$^{-1}$  &  $p_{\mathrm{binom}}$ & $p_{{t \hbox{-} \mathrm {test}}}$\\
\hline
1  & 53561.011--53623.827 &  62.817 & 252.6 & 4.5~~~~& $0.011$ & $8.70\times 10^{-5}$\\
2  & 54272.986--54358.503 &  85.516 & 462.2 & 5.7~~~ & $0.047$ & $8.59\times 10^{-4}$\\
3  & 54746.563--54757.818 &  11.255 & $-$103.5 & $-$12.1~~~ & $0.031$ & $0.019$\\
4  & 54975.245--55070.195 &  94.951 & $-3$73.2 & $-$4.1~~~ & $9.16\times 10^{-4}$ & $0.045$\\
5  & 55066.922--55083.660 &  16.738 & 142.8 & 11.2~~~ & $0.031$ & $3.83\times 10^{-3}$\\
6  & 55170.736--55196.715 &  25.979 & 151.3 & 8.1~~~ & $0.011$ & $0.063$\\
7  & 55344.029--55495.613 &  151.584 & 430.5 & 3.1~~~ & $7.57\times 10^{-4}$ & $0.025$\\
8  & 56453.73--56511.647 &  57.911 & 194.4 & 4.4~~~ & $7.81\times 10^{-3}$ & $0.071$\\
9  & 56520.947--56560.438 &  39.490 & 133.8 & 4.5~~~ & $0.055$ & $3.97\times 10^{-3}$\\
10  & 56560.921--56571.471 &  10.550 & $-$44.5 & $-$4.6~~~ & $3.91\times 10^{-3}$ & $3.27\times 10^{-5}$\\
11  & 56888.015--56989.194 &  101.179 & 533.3 & 5.5~~~ & $0.023$ & $5.68\times 10^{-3}$\\
12  & 57284.646--57323.783 &  39.137 & 412.8 & 11.1~~~ & $1.30\times 10^{-3}$ & $8.97\times 10^{-3}$\\
13  & 57570.997--57618.028 &  47.031 & 382.1 & 9.1~~~ & $3.69\times 10^{-3}$ & $3.53\times 10^{-3}$\\
14  & 57660.555--57682.864 &  22.309 & 206.7 & 12.0~~~ & $0.031$ & $0.059$\\
15  & 57687.548--57756.252 &  68.704 & 330.3 & 5.8~~~ & $9.77\times 10^{-4}$ & $4.30\times 10^{-3}$\\
16  & 57912.430--57999.893 &  87.464 & $-$188.5 & $-$2.8~~~ & $0.062$ & $0.029$\\
17  & 58417.826--58434.294 &  16.467 & 252.8 & 18.5~~~ & $7.81\times 10^{-3}$ & $0.014$\\ \hline
\end{tabular}\\
\end{table*}
\renewcommand{\baselinestretch}{1.0}

\begin{figure}
\includegraphics[width=0.6\textwidth, bb=3 10 210 210,clip]{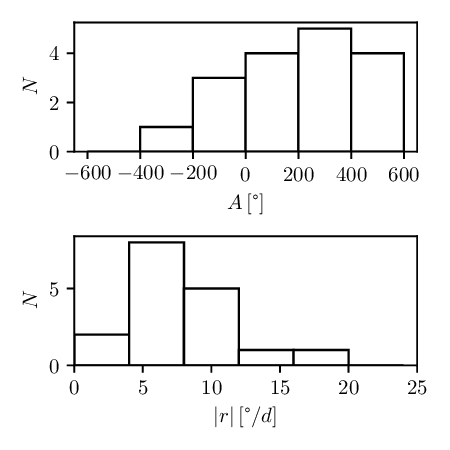}
\caption{Distribution of (a)~amplitudes and (b)~rotation rates of the EVPA for object 3C\,454.3.}
 \label{fig:hist_3c454}
\end{figure}

\begin{figure}
\includegraphics[width=0.6\textwidth, bb=3 10 210 215,clip]{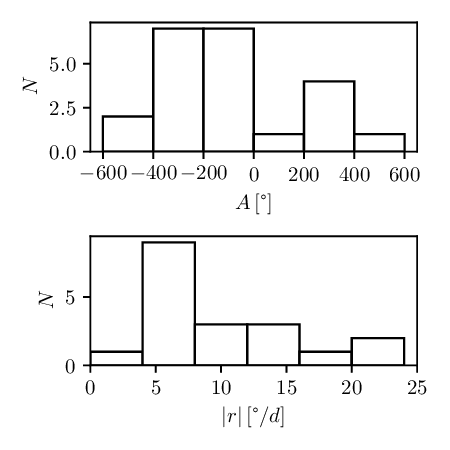}
\caption{Distribution of (a)~amplitudes and (b)~rotation rates of the EVPA for object CTA\,102.}
 \label{fig:hist_cta102}
\end{figure}

\begin{figure}
\includegraphics[width=0.6\textwidth, bb=3 10 210 215,clip]{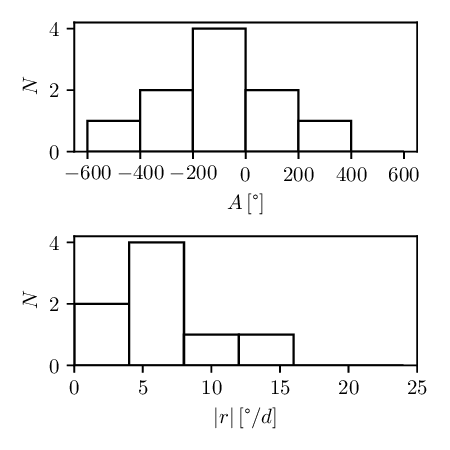}
\caption{Distribution of (a)~amplitudes and (b)~rotation rates of the EVPA for object OT\,081.}
 \label{fig:hist_ot081}
\end{figure}

A total of 51~rotations were identified: 17~rotations for 3C\,454.3, 23~rotations for CTA\,102, and 11~rotations for OT\,081. Currently, this is the most complete sample of known rotations for these objects, obtained as a result of a systematic search. The values of the EVPA for all detected rotations are shown in 
Fig.~\ref{fig:3c454_all_rotations}(3C\,454.3), 
Fig.~\ref{fig:cta102_all_rotations} (CTA\,102) and 
Fig.~\ref{fig:ot081_all_rotations} (OT\,081).  
The rotation parameters (dates, amplitudes, values of average rates) are given in Tables~\ref{tab:rotations3C454}, \ref{tab:rotationsCTA102} and~\ref{tab:rotationsOT081}. The sign of the amplitude determines the direction of rotation (positive is for counterclockwise rotation). For clarity, the amplitudes and average rotation rates are also shown in histograms in Figs.~\ref{fig:hist_3c454}, \ref{fig:hist_cta102} and \ref{fig:hist_ot081}.

\renewcommand{\baselinestretch}{0.75}
\begin{table*}[]
\caption{Rotations parameters of the object CTA\,102, the columns are similar to Table~\ref{tab:rotations3C454}}
\label{tab:rotationsCTA102}
\medskip
\begin{tabular}{c|c|c|r|r|c|c}
\hline
Number  &  MJD & $\Delta \mathrm{MJD}$  &  $A$, deg  & $\bar{r}$, deg day$^{-1}$ &  $p_{\mathrm{binom}}$ & $p_{{t \hbox{-} \mathrm{test}}}$\\
\hline
1  & 54002.357--54074.267 &  71.911 & $-$268.9 & $-$4.9~~~& $0.031$ & $1.70\times 10^{-3}$\\
2  & 54738.397--54776.005 &  37.608 & $-$242.2 & $-$7.3~~~& $3.91\times 10^{-3}$ & $3.47\times 10^{-3}$\\
3  & 55495.169--55515.673 &  20.504 & $-$341.3 & $-$23.5~~~& $0.031$ & $5.87\times 10^{-4}$\\
4  & 55825.572--55831.556 &  5.984 & 346.6 & 63.1~~~& $7.81\times 10^{-3}$ & $0.028$\\
5  & 56190.842--56194.866 &  4.024 & $-$76.3 & $-$23.3~~ & $0.035$ & $0.037$\\
6  & 56215.792--56246.202 &  30.410 & $-$274.9 & $-$9.3~~ & $9.61\times 10^{-3}$ & $1.08\times 10^{-5}$\\
7  & 56485.953--56547.393 &  61.440 & $-$313.6 & $-$5.7~~ & $3.91\times 10^{-3}$ & $0.024$\\
8  & 56575.889--56606.226 &  30.337 & $-$192.8 & $-$7.8~~ & $0.031$ & $0.019$\\
9  & 56838.232--56962.820 &  124.587 & $-$632.6 & $-$5.4~~ & $0.061$ & $0.026$\\
10  & 56917.906--56962.820 &  44.913 & $-$543.3 & $-$12.7~~ & $6.56\times 10^{-4}$ & $0.030$\\
11  & 57243.460--57325.784 &  82.324 & $-$417.9 & $-$5.5~~ & $0.032$ & $0.012$\\
12  & 57564.960--57585.968 &  21.008 & $-$167.8 & $-$9.1~~ & $0.020$ & $2.71\times 10^{-3}$\\
13  & 57585.968--57604.989 &  19.021 & $-$107.9 & $-$7.7~~ & $0.031$ & $7.43\times 10^{-4}$\\
14  & 57618.009--57625.012 &  7.003 & $-$144.3 & $-$24.1~~ & $0.031$ & $0.016$\\
15  & 57635.967--57646.911 &  10.944 & $-$101.7 & $-$11.3~~ & $0.016$ & $3.17\times 10^{-3}$\\
16  & 57691.618--57693.644 &  2.026 & 42.7 & 24.6~~ & $0.016$ & $0.036$\\
17  & 58080.137--58150.148 &  70.011 & $-$324.7 & $-$5.1~~ & $7.25\times 10^{-5}$ & $0.018$\\
18  & 58298.467--58320.489 &  22.022 & 256.7 & 14.1~~ & $0.016$ & $0.016$\\
19  & 58353.982--58371.597 &  17.615 & 290.8 & 18.4~~ & $0.016$ & $0.059$\\
20  & 58430.155--58436.943 &  6.789 & $-$179.7 & $-$33.2~~ & $0.031$ & $6.60\times 10^{-3}$\\
21  & 58655.482--58676.988 &  21.506 & $-$208.1 &$ -$13.0~~ & $0.031$ & $8.08\times 10^{-3}$\\
22  & 58744.130--58781.959 &  37.828 & 227.3 & 6.4~~ & $0.035$ & $4.98\times 10^{-3}$\\
23  & 59750.504--59929.098 &  178.594 & 442.3 & 2.7~~ & $0.011$ & $2.15\times 10^{-3}$\\
\hline
\end{tabular}
\end{table*}
\renewcommand{\baselinestretch}{1.0}
\renewcommand{\baselinestretch}{0.75}

\begin{table*}[]
\caption{Rotations parameters of the object OT\,081, the columns are similar to Table~\ref{tab:rotations3C454}}
\label{tab:rotationsOT081}
\medskip
\begin{tabular}{c|c|c|r|r|c|c}
\hline
Number  &  MJD & $\Delta \mathrm{MJD}$  &  $A $, deg  & $\bar{r}$, deg\,day$^{-1}$ &  $p_{\mathrm{binom}}$ & $p_{{t \hbox{-} \mathrm{test}}}$\\
\hline
1  & 55413.364--55433.347 &  19.983 & $-$80.2 & $-$5.8~~ & $0.020$ & $6.69\times 10^{-3}$\\
2  & 56066.445--56080.470 &  14.026 & 67.2 & 7.4~~ & $0.016$ & $0.032$\\
3  & 56138.370--56171.333 &  32.963 & $-$239.8 & $-$10.0~~ & $0.016$ & $1.36\times 10^{-3}$\\
4  & 56522.806--56535.281 &  12.476 & $-$394.5 & $-$35.1~~ & $3.91\times 10^{-3}$ & $4.55\times 10^{-4}$\\
5  & 56897.216--56916.755 &  19.539 & $-$426.5 & $-$24.8~~ & $9.77\times 10^{-4}$ & $3.07\times 10^{-3}$\\
6  & 56922.761--56931.442 &  8.680 & 238.7 & 33.1~~ & $0.031$ & $2.13\times 10^{-3}$\\
7  & 57103.012--57241.859 &  138.848 & $-$659.4 & $-$5.0~~ & $3.05\times 10^{-5}$ & $0.010$\\
8  & 57243.833--57255.821 &  11.987 & $-$153.4 & $-$15.7~~ & $0.016$ & $4.15\times 10^{-4}$\\
9  & 57577.880--57592.899 &  15.019 & 30.4 & 2.3~~ & $0.031$ & $0.022$\\
10  & 57591.399--57604.847 &  13.448 & $-$34.5 & $-$2.8~~ & $7.81\times 10^{-3}$ & $3.01\times 10^{-3}$\\
11  & 59034.403--59080.562 &  46.159 & $-$161.0 & $-$4.9~~ & $0.031$ & $0.045$\\
\hline
\end{tabular}
\end{table*}
\renewcommand{\baselinestretch}{1.0}

The average frequency of observation of rotations for these objects in the observer’s system is 1.05~(3C\,454.3), 1.31~(CTA\,102) and 0.93~(OT\,081) events per year. Time in the jet system is related to time in the observer system as follows:
$$\Delta T_{\mathrm{jet}}=\Delta T_{\mathrm{obs}}\cfrac{\delta}{1+z},  $$
where $\delta$ is the jet Doppler factor, and $z$ is the redshift. Using Doppler factor estimates from \cite{Weaver2022}, the following average rotation frequencies in the jet system are calculated:0.068 (3C\,454.3), 0.065 (CTA\,102) and 0.047 (OT\,081) events per year.

When compared with data from other researchers using other instruments \citep[for example,][]{Blinov2015,Ryosuke2016}, it can be noted that our rotations parameters are consistent with those previously published. Thus, the duration of the shortest rotations in the Robopol program is 5–7 days \citep{Blinov2015}, while the minimum amplitude is at least 90$^\circ$ due to criteria limitations. The duration of some of the rotations we discovered exceeded 100~days, while, according to RoboPol, the maximum duration is 90~days \citep{Blinov2016b}, and according to the KANATA telescope, long-term rotations of more than 100 days are also visible \citep{Ryosuke2016}. Some models predict faster and shorter rotations \citep{Hosking2020}, and such rotations are indeed detected \citep{Magic2018}, but this occurs sporadically when such an event coincides with a dense observation campaign. When systematically searching for rotations in long-term observational programs, such a density of observations cannot be achieved, and the upper limit of the observed rotation rate is determined by the average frequency of observations \citep{Kiehlmann2021}. 

The significant number of rotations detected in this work allows us to carry out a statistical analysis. In particular, the histograms for objects 3C\,454.3 and CTA\,102 show a clear asymmetry in the distribution of rotation amplitudes relative to zero. For object 3C\,454.3, 14 of the 17 detected rotations occur counterclockwise, and for CTA\,102, 17 of the 23 detected
rotations occur clockwise. The probabilities of such (or greater) asymmetry with a random distribution of rotation directions are 0.025 and 0.017, respectively. At the same time, although the object OT\,081 has asymmetry in the direction of rotation (8 out of 11 rotations occur clockwise), its significance is lower: the probability of such or greater asymmetry is 0.113.

The presence of the EVPA rotations can be explained by the spiral structure of the magnetic field in the jet (for example, \cite{Marscher2008}, model of a shock wave traveling along the jet). In this case, the observed dominant direction of rotations reflects the global structure of the magnetic field, which is related to the direction of rotation of the black hole or accretion disk \citep{Semenov2004}.

It is assumed that in the acceleration and collimation zone closer to the black hole, the magnetic field should be twisted into a tighter spiral \citep{Vlahakis2006}, and with distance, on parsec scales, the degree of twist decreases. Thus, the different rotation rates we observe in the same object may be an indication of the position of the emission region in the jet. On the other hand, observations show \citep{Weaver2022} that for an individual object, the apparent velocity of the radio components in the jet can differ significantly. This will also be reflected in the rotation rate if the rotations are associated with the movement of superluminal components in the jet. In addition, geometric effects associated with different viewing angles of individual sections of the curved jet may have an influence, through a change in the Doppler factor \citep{Raiteri2017}.

The existence of rotations occured in the opposite direction from the dominant may indicate the simultaneous action of the mechanism of random walk of the polarization vector as a result of turbulent movements in the jet. In this model, the emission from individual turbulent cells of the jet, in which the field is assumed to be uniform, adds up to the total emission. The observed polarization is determined by the sum of the Stokes parameters from individual cells \citep{Marscher2014}. If a stochastic mechanism takes place, then it should produce rotations in both directions with equal probability. Therefore, one cannot expect all rotations in the dominant direction to reflect the global structure of the jet’s magnetic field, since the set of such rotations also contains stochastic rotations. However, one can expect differences in the statistical characteristics of rotations occured in different directions. For example, the probability of stochastic rotation decreases with increasing amplitude, since it requires long-term random alignment of the magnetic field in disconnected turbulent cells. For the objects studied in this work, the average amplitude of rotations in the dominant direction ($\left< \Delta \chi_{\mathrm{rot}} \right>$) and in the opposite direction ($\left< \Delta \chi_{\mathrm{counter}} \right>$) is given in Table 4. The table shows that for two out of three objects, the average amplitudes of rotations in the dominant direction significantly exceed the aver age amplitudes of rotations in the opposite direction, which may indicate that the contribution of random walks is insignificant. However, to confirm this, a study of a larger sample of objects is required.

\begin{table}[]
 \renewcommand{\baselinestretch}{0.8}
\caption{Average rotation amplitudes in the dominant direction (2nd column) and the opposite direction (3rd column)} \medskip
    \label{tab:mean_amplitudes}
       \begin{tabular}{c|c|c}
    \hline
         Object    &  $\left< \Delta \chi_{\mathrm{rot}} \right> $, deg  &  $\left< \Delta \chi_{\mathrm{counter}} \right> $, deg\\
         \hline
 (1)  &  (2) &  (3)\\
         \hline
         3C\,454.3  &   $298 \pm 37$      &    $177 \pm 72$ \\
         CTA\,102   &   $271 \pm 37$      &    $267 \pm 55$ \\
         OT\,081    &   $269 \pm 74$      &    $110 \pm 64$ \\
         \hline
    \end{tabular}
    
\end{table}
\renewcommand{\baselinestretch}{1}

\section{CONCLUSIONS}
\label{sec:conclusions}

In this work, a new approach to identifying rotations of the EVPA was proposed and implemented. The method is based on two statistical criteria that
allow assessing the reliability of the found rotations, that is the likelihood of their random occurrence. Compared to previous work, the new method has greater flexibility, allowing us to find not only rotations with a strictly monotonic variation of the EVPA but also rotations in which the EVPA briefly deviates from monotonicity while maintaining the average direction
of rotation. In addition, the method does not have a restriction on the minimum amplitude of rotation; the statistical significance of rotations is determined by the number and accuracy of observations.

The proposed method was tested in numerical experiments on artificial data and demonstrated robustness to observational errors, both in terms of detecting false rotations due to $\chi$ random walk within the errors, and in terms of detecting true noisy rotations. Application of the method to three blazars (3C\,454.3, CTA\,102, OT\,081) made it possible to detect 51~events of significant EVPA rotations: this is the largest sample of such events currently published in the literature. In the future, we plan to apply the described method to a larger sample of galaxies with active nuclei to systematically study the rotations parameters and compare them to the behavior of objects in the optical and other regions of the spectrum

\section*{FUNDING}
The study was supported by a grant from the
Russian Science Foundation \textnumero~23-22-00121, {\url{https://rscf.ru/project/23-22-00121/}}.

\begin{acknowledgments}
The authors thank the anonymous reviewers for
their comments, which helped to significantly improve this work.
\end{acknowledgments} 

\section*{CONFLICT OF INTEREST}
The authors of this work declare that they have no conflicts of interest.

\end{document}